**Revisiting Apophis 2029 approach to Earth (staying on shoulders of NASA's experts)** ***or***

**Can we be sure in almost ricocheting fly-by of Apophis on 13 of April 2029 near the Earth?**

**Sergey Ershkov***, Affiliation[1]: Plekhanov Russian University of Economics,

Scopus number 60030998, e-mail: sergej-ershkov@yandex.ru

Affiliation[2]: Sternberg Astronomical Institute, M.V. Lomonosov's Moscow State University, 13 Universitetskij prospect, Moscow 119992, Russia,

**Dmytro Leshchenko**, Odessa State Academy of Civil Engineering

and Architecture, Odessa, Ukraine, e-mail: leshchenko_d@ukr.net

**Abstract**

The main idea of this challenging research is to revisit the solar-centric dynamics of Earth around the Sun in analysis of its position on 13 April 2029 close to asteroid Apophis which is supposed to be moving in fly-by near the Earth on its orbit. As of now, we can be sure that trajectory of Apophis is well-known with respect to the center of Sun. Also, NASA experts calculated that relative distance between center of Earth and Apophis should be less than 38 thousands of kilometers during closest Apophis approach to the Earth. But the reasonable question is: will the center of Earth be at the predicted position at the beginning of April 2029? The matter is that NASA solving procedure disregards influence of Milankovich cycles to the orbit of Earth but alternative concept suggests another solution (with additional quasi-periodic deviation from their solution, proportional to square of eccentricity of Earth's orbit around the Sun equals to ~ 0.017). So, possible perturbation of Earth orbit is likely to be proportional to $(0.017)^2$ ~ 0.03% from 1 a.e. or ~ 43 200 km which could be compared with gap between Earth and Apophis during closest Apophis approach to Earth in April 2029.

## 1. Introduction, basic assumptions for finding the relative distance between center of Earth and Apophis.

A lot of meaningful attempts have been made during last 18 years in the field of Celestial mechanics for theoretical describing or prediction the proper dynamics of Apophis with respect to the Earth [1-6]. The main efforts have been made by NASA experts [5] and their scientific partners (with world-wide significant results in the field of research under the current investigation). As of now, we can be sure with their very professional help that trajectory of Apophis is well-known with respect to the center of Sun. Also, they calculated that mutual distance between center of Earth and Apophis should be less than 38 thousands of kilometers during closest Apophis fly-by near the Earth on 13 of April 2029. But the reasonable question arises immediately: will center of Earth be at the predicted relative position at the beginning of April 2029? The matter is that experts of NASA are sure regarding that concept of Milankovich cycles can be ignored for their solving procedure at all [5], but the modern research [7] suggests another solution based on keen analysis of data in [8-9] according to recently reported method [10] (with additional quasi-periodic deviation from their solution, proportional to square of eccentricity of Earth's orbit around the Sun equals to ~ 0.017). So, possible perturbation of Earth orbit is likely to be proportional to $(0.017)^2 \sim 0.03\%$ from 1 a.e. or approx. ~ 43 200 km.

In most optimistic scenario (excluding direct impact of Apophis to the Earth's surface), Apophis will pass over the surface of Earth on relative distance circa 0.5 thousands of kilometers. If we take also into account that preliminary trajectory of Apophis will pass overby South Ural located on the territory of Russian Federation Fig.1, this will mean possible interaction of Apophis with upper atmosphere with sufficient density which would appear to cause its explosion like Tunguska event in 1908. The last scenario should be evaluated as catastrophic for the region of South Ural and Kazahstan, albeit we can not exclude a simple scenario of generating a shock wave by explosion (without extra-heat impact onto the local territories within the abovementioned region as a result of asteroid's explosion).

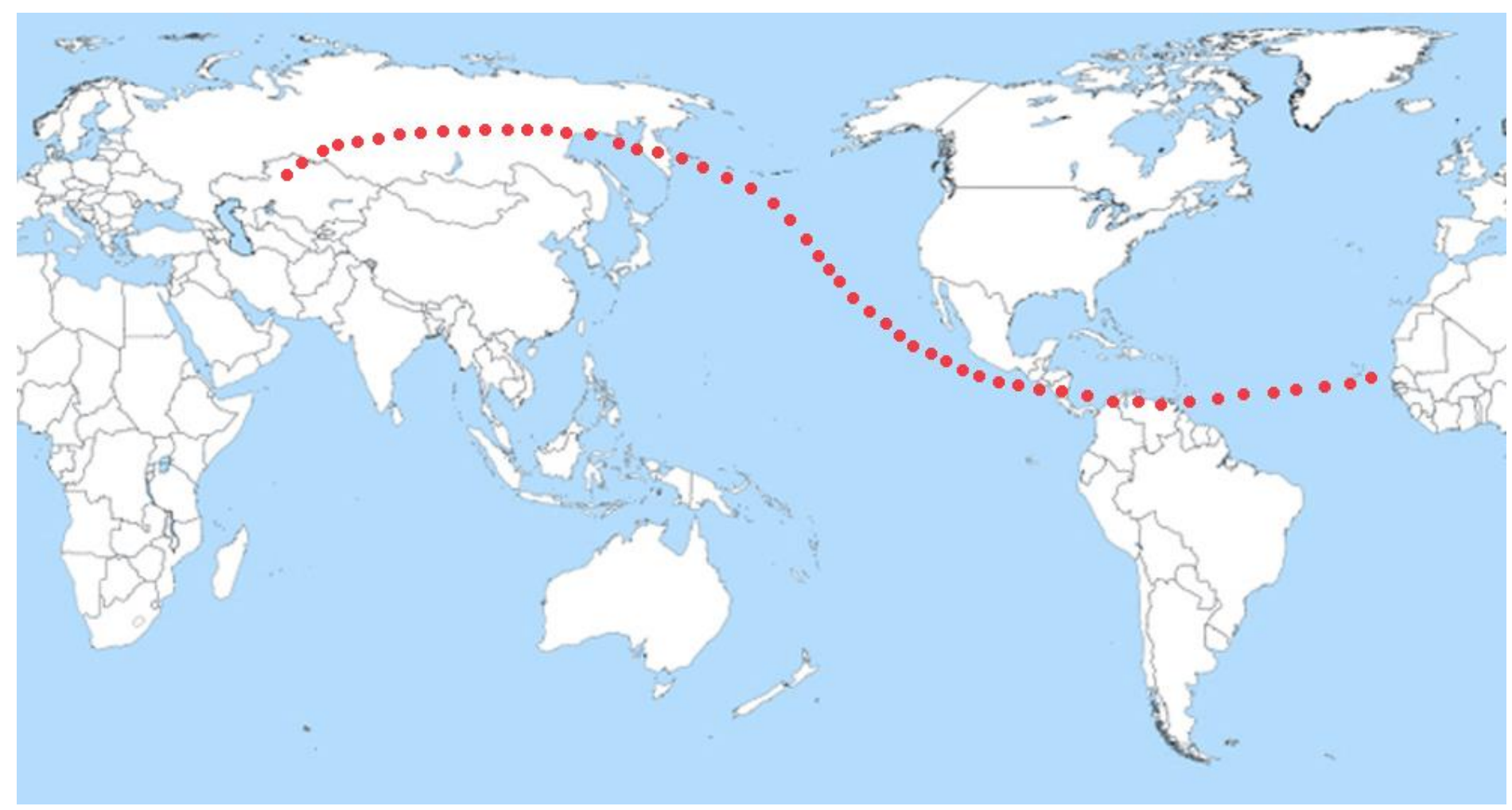

Fig.1. Schematically presented Apophis Path of Risk. The initial position is region of South Ural and Kazahstan.

## 2. Estimation of possible additional contribution to the motion of Earth around the Sun [7].

The main idea of the current research is to revisit the solar-centric dynamics of Earth around the Sun in analysis of its position on 13 April 2029 close to asteroid Apophis fly-by approach to the Earth on its orbit.

First, we should note that a lot of close approaches of NEO to Earth have been noticed during only last year by authors of this research (Table 1), we used on-line service [11]. The Earth planetary system (along with satellites on their orbits around the Earth) appears to definitely be atacked by sufficiently large asteroid in the future. Nevertheless, there are (unpredictable) asteroids in Solar system which would may come to Earth from the side of Sun (this direction is not used for astronomic observations) like as it has happened in Chelyabinsk event in 2013.

Table 1. Recent close approaches of NEO to the Earth.

| **Number, in the order of detecting [11]**<br><br>**with date**<br><br>**(if any)** | **NEO Earth**<br><br>**(close approaches to the Earth)** | **Last solution date**<br><br>**(author)** | **Approx. distance to Earth,**<br><br>**MOID (in au)** |
|---|---|---|---|
| ***Asteroids*** **(*size > 10 m*)** | | | |
| 1.<br><br>(2004-Jun-19) | **99942 Apophis (2004 MN4)**<br>**Classification:** Aten [NEO, PHA]<br><br>**SPKID:** 2099942 | 2021-Jun-29<br>11:09:44<br><br>(**Davide Farnocchia**) | 0.000244177 au |
| 2. | **(2011 ES4)**<br>**Classification:** Apollo [NEO]<br><br>**SPKID:** 3558279 | 2021-Apr-15<br>04:52:34<br><br>(Otto Matic) | 0.000775033 au |
| 3. | **(2020 PY2)**<br>**Classification:** Apollo [NEO]<br><br>**SPKID:** 54051050 | 2021-Jul-07<br>19:03:45<br><br>(Otto Matic) | 0.00126387 au |
| 4. | **(2020 SW)**<br>**Classification:** Aten [NEO]<br><br>**SPKID:** 54054455 | 2021-Apr-15<br>23:02:42<br><br>(Otto Matic) | 7.30264E-7 au |
| 5.<br><br>(2020-Oct-15) | **(2020 UE)**<br>**Classification:** Apollo [NEO]<br><br>**SPKID:** 54073345 | 2021-Oct-15<br>06:31:21<br><br>(Otto Matic) | 0. 00101049 au |

| 6.<br>(2020-Oct-16) | **(2020 TE6)**<br>**Classification:** Apollo [NEO]<br>**SPKID:** 54065913 | 2021-Oct-15<br>06:31:18<br>(Otto Matic) | 0.000324752 au |
| --- | --- | --- | --- |
| 7.<br>(2020-Oct-17) | **(2020 TF6)**<br>**Classification:** Apollo [NEO]<br>**SPKID:** 54065914 | 2021-Oct-15<br>06:31:21<br>(Otto Matic) | 0.000872265 au |
| 8.<br>(2020-Nov-30) | **(2020 VZ6)**<br>**Classification:** Apollo [NEO]<br>**SPKID:** 54096685 | 2021-Apr-15<br>23:17:14<br>(Otto Matic) | 0.00191305 au |
| 9.<br>(2020-Dec-12) | **(2018 BA3)**<br>**Classification:** Apollo [NEO]<br>**SPKID:** 3797848 | 2021-Sep-03<br>05:49:38<br>(Otto Matic) | 0.0022614 au |
| 10. | **(2020 VY1)**<br>**Classification:** Apollo [NEO]<br>**SPKID:** 54087660 | 2021-Apr-15<br>23:16:49<br>(Otto Matic) | 0.0315058 au |
| 11.<br>(2020-Dec-16) | **(2020 XX3)**<br>**Classification:** Apollo [NEO]<br>**SPKID:** 54099638 | 2021-Apr-15<br>23:19:57<br>(Otto Matic) | 0.000126671 au |
| 12.<br>(2020-Dec-16) | **(2020 XF4)**<br>**Classification:** Apollo [NEO]<br>**SPKID:** 54099647 | 2021-Jun-28<br>06:24:36<br>(Otto Matic) | 0.00116043 au |

| 13.<br><br>(2021-Feb-06) | **(2021 CO)**<br>**Classification:** Apollo [NEO]<br><br>**SPKID:** 54110051 | 2021-Nov-25<br>04:54:21<br><br>(Otto Matic) | 0.00146935 au |
|---|---|---|---|
| 14.<br><br>(2021-Feb-14) | **(2021 CS6)**<br>**Classification:** Apollo [NEO]<br><br>**SPKID:** 54117569 | 2021-Apr-15<br>23:25:22<br><br>(Otto Matic) | 0.000215766 au |
| 15.<br><br>(2021-Mar-12) | **(2021 EQ3)**<br>**Classification:** Apollo [NEO]<br><br>**SPKID:** 54131350 | 2021-Apr-15<br>23:26:35<br><br>(Otto Matic) | 0.00136426 au |
| 16.<br><br>(2021-Mar-19) | **(2021 FH)**<br>**Classification:** Apollo [NEO]<br><br>**SPKID:** 54132049 | 2021-May-31<br>06:00:21<br><br>(Otto Matic) | 0.000153384 au |
| 17.<br><br>(2021-Apr-08) | **(2021 GT3)**<br>**Classification:** Apollo [NEO]<br><br>**SPKID:** 54135430 | 2021-Apr-15<br>23:27:06<br><br>(Otto Matic) | 0.00142861 au |
| 18.<br><br>(2021-Apr-09) | **(2021 GW4)**<br>**Classification:** Apollo [NEO]<br><br>**SPKID:** 54135784 | 2021-Apr-15<br>07:01:18<br><br>(Otto Matic) | 8.4254E-5 au |
| 19.<br><br>(2021-Apr-15) | **(2021 GF10)**<br>**Classification:** Apollo [NEO]<br><br>**SPKID:** 54137446 | 2021-Apr-21<br>06:22:52<br><br>(Otto Matic) | 0.000207837 au |
| 20.<br>(2021-Apr-19) | **(2021 HN)**<br>**Classification:** Apollo [NEO]<br>**SPKID:** 54138695 | 2021-Apr-20 06:22:15<br>(Otto Matic) | 0.00152674 au |
| 21.<br>(2021-May-04) | **(2021 JV)**<br>**Classification:** Apollo [NEO]<br>**SPKID:** 54143222 | 2021-May-15 06:33:21<br>(Otto Matic) | 0.000616189 au |
| 22.<br>(2021-May-06) | **(2021 JS1)**<br>**Classification:** Apollo [NEO]<br>**SPKID:** 54143557 | 2021-May-07 05:55:33<br>(Otto Matic) | 5.26962E-5 au |
| 23.<br>(2021-May-14) | **(2021 JU6)**<br>**Classification:** Apollo [NEO]<br>**SPKID:** 54145471 | 2021-May-15 06:33:27<br>(Otto Matic) | 0.000747435 au |
| 24.<br>(2021-May-31) | **(2021 KN2)**<br>**Classification:** Apollo [NEO]<br>**SPKID:** 54149826 | 2021-Jun-01 05:50:56<br>(Otto Matic) | 0.000708473 au |
| 25.<br>(2021-Jul-02) | **(2021 NA)**<br>**Classification:** Apollo [NEO]<br>**SPKID:** 54164554 | 2021-Jul-03 06:40:28<br>(Otto Matic) | 0.000260047 au |
| 26. | **(2021 RR5)**<br>**Classification:** Apollo [NEO]<br>**SPKID:** 54194342 | 2021-Sep-23 05:50:54<br>(Otto Matic) | 0.000228105 au |

| 27.<br><br>(2021-Sep-16) | **(2021 SG)**<br>**Undetected previously** [NEO]<br><br>Average size: 68 meters | **Undetected,**<br><br>**came from Sun**<br><br>(2021-Sep-17) | ~ 0.5 LD<br><br>(LD is the distance from Earth to the Moon) |
|---|---|---|---|
| 28.<br><br>(2021-Oct-15) | **(2021 TG14)**<br>**Classification:** Apollo [NEO]<br><br>**SPKID:** 54209303 | 2021-Oct-20 05:51:17<br><br>(Otto Matic) | 0.00160166 au |
| 29.<br><br>(2021-Nov-09) | **(2021 VU4)**<br>**Classification:** Apollo [NEO]<br><br>**SPKID:** 54217191 | 2021-Nov-10 08:24:33<br><br>(Otto Matic) | 0.000165451 au |
| 30.<br><br>(2021-Nov-11) | **(2021 VC7)**<br>**Classification:** Apollo [NEO]<br><br>**SPKID:** 54217977 | 2021-Nov-17 04:49:29<br><br>(Otto Matic) | 0.00137678 au |
| 31.<br><br>(2021-Dec-10) | **(2021 XV4)**<br>**Classification:** Apollo [NEO]<br><br>**SPKID:** 54229476 | 2021-Dec-09 18:12:02<br><br>(Otto Matic) | 0.000465116 au |
| 32.<br><br>(2021-Dec-15) | **(2021 XC6)**<br>**Classification:** Aten [NEO]<br><br>**SPKID:** 54230077 | 2021-Dec-29 04:50:27<br><br>(Otto Matic) | 0.00125937 au |
| 33.<br><br>(2021-Dec-29) | **(2021 YK)**<br>**Classification:** Apollo [NEO]<br><br>**SPKID:** 54231688 | 2022-Jan-02 04:51:19<br><br>(Otto Matic) | 0.000959231 au |

| 34. | **137108 (1999 AN10)**<br>**Classification:** Apollo [NEO, PHA]<br>**SPKID:** 2137108<br><br>Size: 700 meters – 1.6 km | 2021-Dec-31 04:52:25<br><br>(Otto Matic) | 0.00157691 au |
|---|---|---|---|

As we can see from Table 1, it is very likely for Earth to meet one crucial attack of asteroid from *Aten* or *Apollo* families of NEO in the nearest future.

In the current research, we will restrict ourselves in presenting a new updating in formula related to Earth's solar-centric distance to Sun according to the ansatz presented in [7]. As for the complete introduction to the problem under consideration, we recommend seminal articles [7] and [10], where a significant theoretical explanations have been made as well as all the difficulties for such a kind of Earth's solar-centric motion are considered in details.

With respect to Apophis 2029 future event, formula (1) was presented in [7] for solar-centric distance $r$ (Earth from Sun), taking into account the solar-terrestrial interactions via various types of seasonal irradiation processes influencing to orbit of Earth (including long-term Milankovich cycles)

$$\left(\frac{a_p}{r}\right)^2 = 1 + 2e\cos f + e^2(\cos^2 f + 2) + 4e^3\cos f + O(e^4)\,, \quad \Rightarrow$$

$$\Rightarrow \quad a_p \cong (1 + e\cos f + e^2 + O(e^3))\,r \quad \Rightarrow \quad r \cong (1 - e\cos f - e^2\sin^2 f)\,a_p \qquad (1)$$

where $f$ is the true anomaly, $a_p$ is the current semimajor axis of the Earth's orbit, $e$ is the eccentricity of it's orbit (which is also variable in general case [12-13], average eccentricity of Earth's orbit around the Sun equals to ~ 0.0167). Whereas for the first (ER2BP) approximation for solar-centric distance of celestial body from the Sun we can use formula (2) of classic Kepler's motion:

$$r_A = \frac{a_A}{1 + e_A\cos f_A} \quad \Rightarrow \quad r_A \cong (1 - e_A\cos f_A + \frac{1}{8}e_A{}^2\cos^2 f_A)\,a_A \qquad (2)$$

where (here above and in all formulae below) index $A$ refers to all designations of the case of Apophis for the true anomaly, semimajor axis and eccentricity.

We can conclude from (1) that, *videlicet*, there exist a possible perturbations of Earth orbit proportional to $(0.017)^2$ ~ 0.03% from 1 a.e. (or approx. ~ 43 200 km) with respect to the NASA's solution which were not used in their solving procedure or numerical algorithm for calculations of solar-centric position of Earth on 13 of April 2029, Fig.2.

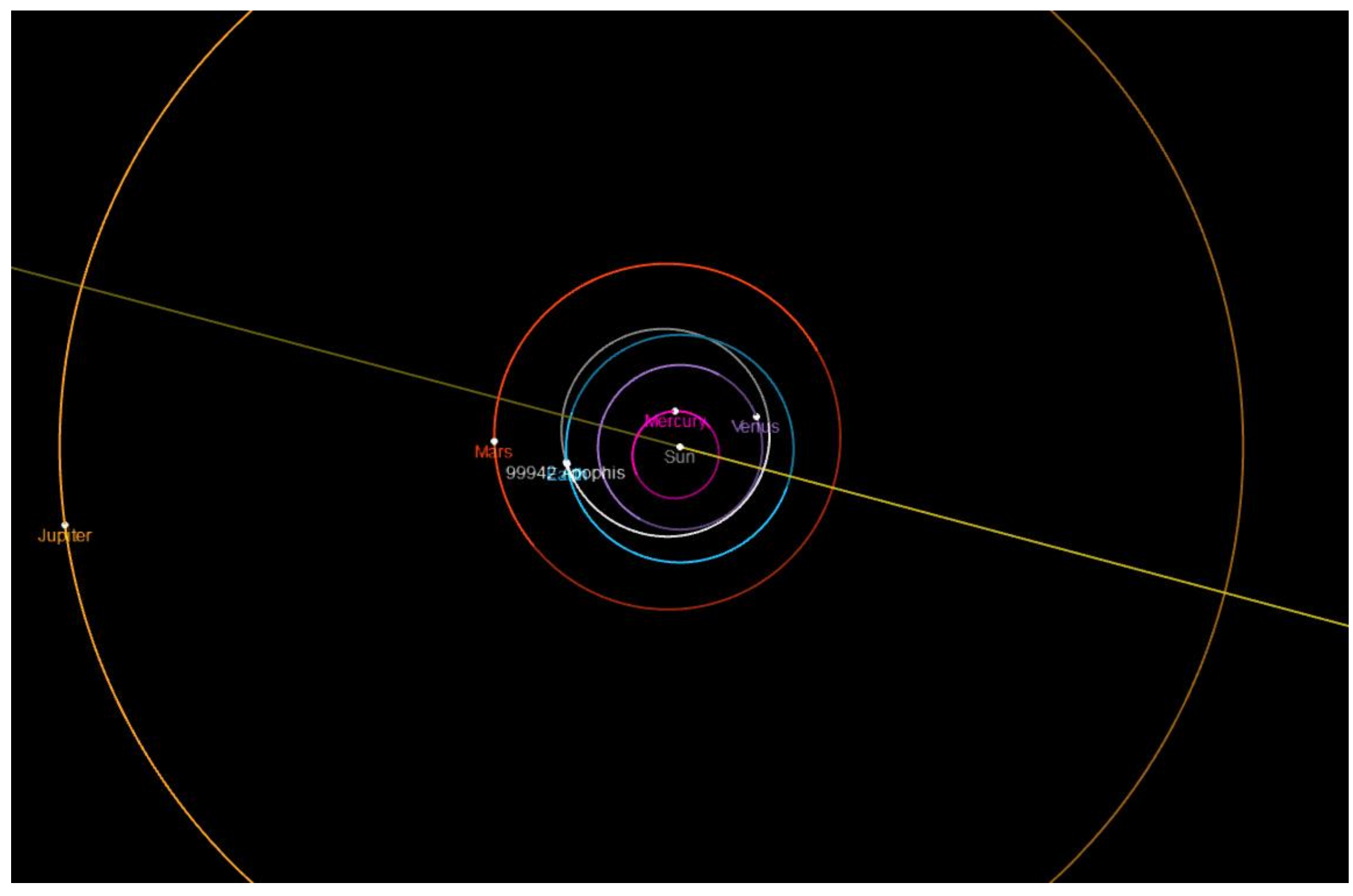

Fig.2. Schematically presented Apophis Close Approach to the Earth according to NASA solution (see [9], [11]).

## 3. Discussion & conclusion.

We introduce here in the current research the revisiting of approach for calculations of solar-centric position of Earth on its orbit in motion around the Sun. A quasi-periodic perturbations for such motion stem from various types of seasonal solar activity (including long-term Milankovich cycles).

As reported in [7], there exist a possible perturbations of Earth orbit proportional to $(0.017)^2$ ~ 0.03% from 1 a.e. (or approx. ~ 43 200 km) with respect to the NASA's solution which were used in their solving procedure or numerical algorithm for calculations of solar-centric position of Earth on 13 of April 2029 (they disagree to the statement that solar activity might have been influencing on orbital motion of Earth during previous middle-time period or, moreover, even 1 year). Based on results reported in [7], we can not agree with such the conservative position and should note that circa 4.5 millions of people will be still living in the region of South Ural and Kazahstan in 2029 year, so it is critically important to exclude ambiguity from the initial assumptions for solving procedure used to obtain the approximate relative distance between center of Earth and Apophis during closest Apophis fly-by near the Earth on 13 of April 2029.

For example, if perturbations of Earth's orbit would be enough to shift its postion closer to the Apophis during assumed event on 13 of April 2029, this mean that trajectory of Apophis will be passing over by the dense near-surface layers of atmosphere of the Earth which may cause additional aerodynamical torques during such motion which appear to be additional crucial driving factors with respect to the dynamics of Apophis during its fly-by over the Earth surface. This may even cause the extra termo-explosion like Tunguska event in 1913. As a result, it would mean a catastrophic event for people living in the region of South Ural, Kazahstan republic or Far East region of Russian Federation.

Referring to the rationale for the aforementioned hypothesis of updating the solving procedure for Earth's orbit, we can formulate the main suggestion as follows (and its significance in the area of future practical using in Astronautica science field within the limits of what is necessary to indicate an actual problem to which the results of the current investigations can be applied): Milankovich cycles and other types of seasonal solar activity due to irradiation from Sun should be taken into consideration as important factor influencing on orbital motion of Earth when it moves around the Sun.

It is worth to list below the additional factors which may influence on the motion of Apophis during its fly-by close to the Earth surface:

1) If trajectory of Apophis will be passing over by the dense near-surface layers of atmosphere of the Earth, it is need to take into account the additional aerodynamical torques [14]. Also (under assumption above), surrounding it's surface plasma will may slightly deviate Apophis (in quasi-oscillating or spiralling-type motion [15]) from the main trajectory due to interacting with EM-field of Earth (or with EM-field generated by phenomena existing in the atmosphere of the Earth);
2) The Lense-Thirring gravito-magnetic effect of GTR [8], governing the small-scaled orbital precessions induced by the spin angular momentum of the central body, should be taken into consideration which may cause the deviations of trajectory of Apophis in its motion relative to the Earth;
3) Influence of Yarkovsky effect [16-17] on trajectory of Apophis should be updated close to the 2029 year;
4) Effect of cosmic billiard (with help of other small celestial bodies crossing the Solar system within the orbit of Mars), which would may suddenly change trajectory of Apophis, could not be estimated even preliminarly as an example of true occasional events which do happen from time to time. As noted in [3], Apophis's closest predictable encounter will be with asteroid 2001 GQ2 at 0.63 lunar distance in January 2027;
5) Some uncertainties in detecting of Apophis trajectory could come from considering Apophis as material point in equations of motion.

As we can see the list of uncertainties in detecting of trajectory of Apophis, presented in [5], could be enlarged insofar for further improving the accuracy in calculating the proper trajectory during its fly-by close to the Earth surface in 2029.

The last but not least, we should note with respect to illuminating the aim and the main findings of the current research that though NASA's experts consider Milankovich cycles in lon-term orbital dynamics or climate change [18] during thousands of years, they do not take into considerations the aforementioned effect in their calculations of motion for orbits of planets of Solar system (including the Earth) on a time-scale of tens of years [19], see **Appendix A1**. But this could be crucial issue for calculating the event of Apophis's approaching the Earth in their close encounter in 2029, according to recent results reported in [7]. Indeed, according to formula (1) above (with values of true anomaly which are reported in **Appendix A1** on date Apr 13, 2029 for the time-period 18:00-21:00), we obtain estimation of solar-centric distance of Earth from Sun at that day for value of true anomaly which varies from $f$ = 97.93° (on time-moment 18:00) to $f$ = 98.05° (on time-moment 21:00):

$$r\big|_{18:00} \cong (1 - e\cos f - e^2 \sin^2 f)\, a_p \cong 1.00288\, a.e. \qquad (3a)$$

$$r\big|_{21:00} \cong (1 - e\cos f - e^2 \sin^2 f)\, a_p \cong 1.00291\, a.e. \qquad (3b)$$

where in ephemeris of NASA, reported in **Appendix A1** below, value $a_p$ varies from 1.00086172 *a.u.* (on time-moment 18:00) to 1.00086176 *a.u.* (on time-moment 21:00), whereas value $e$ varies from 0.016572 (on time-moment 18:00) to 0.016572 0.016561 (on time-moment 21:00), respectively.

Furthermore, instead of formula (2) for first approximation, we will use formula (1) above (with values of true anomaly which are reported in **Appendix A2** on the date Apr 13, 2029 for the time-period 18:00-21:00) to obtain estimation of solar-centric distance of Apophis from Sun at that day for value of true anomaly which varies from $f = 236°$ (on time-moment 18:00) to $f = 249.5°$ (on time-moment 21:00):

$$r_A\Big|_{18:00} \cong (1 - e_A \cos f_A - e_A{}^2 \sin^2 f_A) a_A \cong 1.00720\, a.e. \qquad (4a)$$

$$r_A\Big|_{21:00} \cong (1 - e_A \cos f_A - e_A{}^2 \sin^2 f_A) a_A \cong 1.00646\, a.e. \qquad (4b)$$

where in ephemeris of NASA, reported in **Appendix A2** below, value $a_A$ varies from 0.92656602 *a.u.* (on time-moment 18:00) to 0.97289097 *a.u.* (on time-moment 21:00), whereas value $e_A$ varies from 0.2096497 (on time-moment 18:00) to 0.2219692 (on time-moment 21:00), respectively.

As we can see, difference of {(4*a*) – (3*a*)} equals to 4.32E-03 *a.u.* ≅ 646 263 km (which exceeds distance to the Moon), whereas difference of {(4*b*) – (3*b*)} gives value 3.55E-03 *a.u.* ≅ 531 073 km (which also exceeds distance to the Moon). Thus, the obtained results exclude pessimistic scenario (it means that Apophis will successfully fly-by far from the Earth at the chosen time-period from 18:00 to 21:00 on 13 of April 2029), but there may exist another time-period of close approach which should be calculated additionally.

It is of keen interest to calculate the real date of maximal close approach of Apophis to the Earth in the time-step of the aforementioned calculating scheme.

Using formula (1) above (with value of true anomaly which is reported in **Appendix A1** on date Apr **14**, 2029 for the time-moment 00:00), let us obtain estimation of solar-centric distance of Earth from Sun at that day for value of true anomaly $f = 98.18°$ (on time-moment 00:00):

$$r\Big|_{(Apr14),\,00:00} \cong (1 - e\cos f - e^2 \sin^2 f) a_p \cong 1.00295\, a.e. \qquad (5)$$

where in ephemeris of NASA, reported in **Appendix A3** below, value $a_P$ equals to 1.00086128 *a.u.* (on time-moment 00:00), whereas value $e$ equals to 0.016551 (on time-moment 00:00), respectively.

Furthermore, instead of formula (2) for first approximation, we will use formula (1) above (with value of true anomaly which is reported in **Appendix A2** on the date Apr **14**, 2029 for the time-moment 00:00) to obtain estimation of solar-centric distance of Apophis from Sun at that day for value of true anomaly $f$ = 283° (on time-moment 00:00):

$$r_A\Big|_{(Apr14),\,00:00} \cong (1 - e_A \cos f_A - e_A{}^2 \sin^2 f_A)\, a_A \cong 1.0041\, a.e. \qquad (6)$$

where in ephemeris of NASA, reported in **Appendix A4** below, value $a_A$ equals to 1.1032804 *a.u.* (on time-moment 00:00), whereas value $e_A$ equals to 0.21656 (on time-moment 00:00), respectively.

As we can see, difference of {(6) – (5)} gives value 2.54E-03 *a.u.* ≅ 379 915 km (which could be compared with the distance to the Moon). Thus, the obtained results also exclude pessimistic scenario (it means that Apophis will successfully fly-by far from the Earth at the chosen time-period 00:00 on 14 of April 2029), albeit we can not exclude Apophis close approach to the Moon.

Also, article [20] should be cited which concerns the problem under consideration.

## Appendix A1 (calculations of ephemeris of Earth's orbit on 2029-Apr-13 with time-step 1 hour, using on-line service NASA [9]).

Let us we calculate ephemeris of Earth's orbit on 2029-Apr-13 with time-step 1 hour, using on-line service NASA [9]:

```
*******************************************************************************
 Revised: April 12, 2021                 Earth                              399

 GEOPHYSICAL PROPERTIES (revised Aug 15, 2018):
  Vol. Mean Radius (km)    = 6371.01+-0.02   Mass x10^24 (kg)= 5.97219+-0.0006
  Equ. radius, km          = 6378.137        Mass layers:
  Polar axis, km           = 6356.752          Atmos         = 5.1   x 10^18 kg
  Flattening               = 1/298.257223563   oceans        = 1.4   x 10^21 kg
  Density, g/cm^3          = 5.51              crust         = 2.6   x 10^22 kg
  J2 (IERS 2010)           = 0.00108262545     mantle        = 4.043 x 10^24 kg
  g_p, m/s^2  (polar)      = 9.8321863685      outer core    = 1.835 x 10^24 kg
  g_e, m/s^2  (equatorial) = 9.7803267715      inner core    = 9.675 x 10^22 kg
  g_o, m/s^2               = 9.82022         Fluid core rad  = 3480 km
  GM, km^3/s^2             = 398600.435436   Inner core rad  = 1215 km
  GM 1-sigma, km^3/s^2     =      0.0014     Escape velocity = 11.186 km/s
  Rot. Rate (rad/s)        = 0.00007292115   Surface area:
  Mean sidereal day, hr    = 23.9344695944     land          = 1.48 x 10^8 km
  Mean solar day 2000.0, s = 86400.002         sea           = 3.62 x 10^8 km
  Mean solar day 1820.0, s = 86400.0         Love no., k2    = 0.299
  Moment of inertia        = 0.3308          Atm. pressure   = 1.0 bar
  Mean temperature, K      = 270             Volume, km^3    = 1.08321 x 10^12
  Mean effect. IR temp, K  = 255             Magnetic moment = 0.61 gauss Rp^3
  Geometric albedo         = 0.367           Vis. mag. V(1,0)= -3.86
  Solar Constant (W/m^2)   = 1367.6 (mean), 1414 (perihelion), 1322 (aphelion)
 HELIOCENTRIC ORBIT CHARACTERISTICS:
  Obliquity to orbit, deg  = 23.4392911  Sidereal orb period  = 1.0000174 y
  Orbital speed, km/s      = 29.79       Sidereal orb period  = 365.25636 d
  Mean daily motion, deg/d = 0.9856474   Hill's sphere radius = 234.9
*******************************************************************************

*******************************************************************************
Ephemeris / WWW_USER Sun Jan  2 06:24:04 2022 Pasadena, USA      / Horizons
*******************************************************************************
Target body name: Earth (399)                     {source: DE441}
Center body name: Sun (10)                        {source: DE441}
Center-site name: BODY CENTER
*******************************************************************************
Start time      : A.D. 2029-Apr-13 00:00:00.0000 TDB
Stop  time      : A.D. 2029-Apr-14 00:00:00.0000 TDB
Step-size       : 60 minutes
*******************************************************************************
Center geodetic : 0.00000000,0.00000000,0.0000000 {E-lon(deg),Lat(deg),Alt(km)}
Center cylindric: 0.00000000,0.00000000,0.0000000 {E-lon(deg),Dxy(km),Dz(km)}
Center radii    : 696000.0 x 696000.0 x 696000.0 k{Equator, meridian, pole}
```

```
Keplerian GM    : 2.9591309705336418E-04 au^3/d^2
Output units    : AU-D, deg, Julian Day Number (Tp)
Output type     : GEOMETRIC osculating elements
Output format   : 10
Reference frame : Ecliptic of J2000.0
*******************************************************************************
JDTDB
   EC     QR    IN
   OM     W     Tp
   N      MA    TA
   A      AD    PR
*******************************************************************************
$$SOE
2462239.500000000 = A.D. 2029-Apr-13 00:00:00.0000 TDB
 EC= 1.663657428868264E-02 QR= 9.842001562574755E-01 IN= 3.965790242761895E-03
 OM= 1.785797518429113E+02 W = 2.871612117832327E+02 Tp=  2462142.673280525487
 N = 9.843525232787116E-01 MA= 9.531162563532909E+01 TA= 9.720585777018148E+01
 A = 1.000850886380641E+00 AD= 1.017501616503806E+00 PR= 3.657226364401454E+02
2462239.541666667 = A.D. 2029-Apr-13 01:00:00.0000 TDB
 EC= 1.663295595067934E-02 QR= 9.842048389688457E-01 IN= 3.982084579234202E-03
 OM= 1.786859964545832E+02 W = 2.870575404403909E+02 Tp=  2462142.675328800920
 N = 9.843509310984041E-01 MA= 9.535046986989933E+01 TA= 9.724414468262933E+01
 A = 1.000851965626258E+00 AD= 1.017499092283671E+00 PR= 3.657232279937888E+02
2462239.583333333 = A.D. 2029-Apr-13 02:00:00.0000 TDB
 EC= 1.662933905823985E-02 QR= 9.842094653376772E-01 IN= 3.998372474309704E-03
 OM= 1.787914461009131E+02 W = 2.869544634894908E+02 Tp=  2462142.677180839237
 N = 9.843494212894219E-01 MA= 9.538951512715212E+01 TA= 9.728263122879312E+01
 A = 1.000852989039874E+00 AD= 1.017496512742071E+00 PR= 3.657237889452179E+02
2462239.625000000 = A.D. 2029-Apr-13 03:00:00.0000 TDB
 EC= 1.662572389038696E-02 QR= 9.842140350246347E-01 IN= 4.014652660537436E-03
 OM= 1.788961037317638E+02 W = 2.868519777110095E+02 Tp=  2462142.678836402483
 N = 9.843479939403421E-01 MA= 9.542876165748480E+01 TA= 9.732131767751672E+01
 A = 1.000853956561019E+00 AD= 1.017493878097405E+00 PR= 3.657243192612412E+02
2462239.666666667 = A.D. 2029-Apr-13 04:00:00.0000 TDB
 EC= 1.662211072685105E-02 QR= 9.842185476918427E-01 IN= 4.030923870499263E-03
 OM= 1.789999723298170E+02 W = 2.867500798617592E+02 Tp=  2462142.680295263883
 N = 9.843466491365190E-01 MA= 9.546820970211226E+01 TA= 9.736020428848887E+01
 A = 1.000854868131431E+00 AD= 1.017491188571020E+00 PR= 3.657248189098793E+02
2462239.708333333 = A.D. 2029-Apr-13 05:00:00.0000 TDB
 EC= 1.661849984806140E-02 QR= 9.842230030029079E-01 IN= 4.047184836926506E-03
 OM= 1.791030549079179E+02 W = 2.866487666776185E+02 Tp=  2462142.681557205506
 N = 9.843453869600687E-01 MA= 9.550785949299296E+01 TA= 9.739929131216840E+01
 A = 1.000855723695066E+00 AD= 1.017488444387223E+00 PR= 3.657252878603716E+02
2462239.750000000 = A.D. 2029-Apr-13 06:00:00.0000 TDB
 EC= 1.661489153513802E-02 QR= 9.842274006229297E-01 IN= 4.063434292796876E-03
 OM= 1.792053545063751E+02 W = 2.865480348762958E+02 Tp=  2462142.682622018270
 N = 9.843442074898602E-01 MA= 9.554771125276427E+01 TA= 9.743857898971989E+01
 A = 1.000856523198101E+00 AD= 1.017485645773273E+00 PR= 3.657257260831784E+02
2462239.791666667 = A.D. 2029-Apr-13 07:00:00.0000 TDB
 EC= 1.661128606988285E-02 QR= 9.842317402185166E-01 IN= 4.079670971357243E-03
 OM= 1.793068741902970E+02 W = 2.864478811600598E+02 Tp=  2462142.683489501476
 N = 9.843431108015105E-01 MA= 9.558776519467682E+01 TA= 9.747806755294768E+01
 A = 1.000857266588946E+00 AD= 1.017482792959376E+00 PR= 3.657261335499841E+02
2462239.833333333 = A.D. 2029-Apr-13 08:00:00.0000 TDB
```

```
 EC= 1.660768373476958E-02 QR= 9.842360214578062E-01 IN= 4.095893606230958E-03
 OM= 1.794076170472570E+02 W = 2.863483022181490E+02 Tp=  2462142.684159466997
 N = 9.843420969673644E-01 MA= 9.562802152251953E+01 TA= 9.751775722422040E+01
 A = 1.000857953818248E+00 AD= 1.017479886178690E+00 PR= 3.657265102337036E+02
2462239.875000000 = A.D. 2029-Apr-13 09:00:00.0000 TDB
 EC= 1.660408481293586E-02 QR= 9.842402440104758E-01 IN= 4.112100931506307E-03
 OM= 1.795075861847629E+02 W = 2.862492947293613E+02 Tp=  2462142.684631731827
 N = 9.843411660564954E-01 MA= 9.566848043056100E+01 TA= 9.755764821641267E+01
 A = 1.000858584838896E+00 AD= 1.017476925667316E+00 PR= 3.657268561084828E+02
2462239.916666667 = A.D. 2029-Apr-13 10:00:00.0000 TDB
 EC= 1.660048958817224E-02 QR= 9.842444075477635E-01 IN= 4.128291681734589E-03
 OM= 1.796067847276854E+02 W = 2.861508553646914E+02 Tp=  2462142.684906125534
 N = 9.843403181346909E-01 MA= 9.570914210348407E+01 TA= 9.759774073283904E+01
 A = 1.000859159606030E+00 AD= 1.017473911664297E+00 PR= 3.657271711497038E+02
2462239.958333333 = A.D. 2029-Apr-13 11:00:00.0000 TDB
 EC= 1.659689834491320E-02 QR= 9.842485117424842E-01 IN= 4.144464592112840E-03
 OM= 1.797052158165079E+02 W = 2.860529807891717E+02 Tp=  2462142.684982487466
 N = 9.843395532644370E-01 MA= 9.575000671629488E+01 TA= 9.763803496716307E+01
 A = 1.000859678077052E+00 AD= 1.017470844411619E+00 PR= 3.657274553339909E+02
2462240.000000000 = A.D. 2029-Apr-13 12:00:00.0000 TDB
 EC= 1.659331136822682E-02 QR= 9.842525562690438E-01 IN= 4.160618398470680E-03
 OM= 1.798028826045958E+02 W = 2.859556676646365E+02 Tp=  2462142.684860663954
 N = 9.843388715049187E-01 MA= 9.579107443429041E+01 TA= 9.767853110336431E+01
 A = 1.000860140211623E+00 AD= 1.017467724154201E+00 PR= 3.657277086392103E+02
2462240.041666667 = A.D. 2029-Apr-13 13:00:00.0000 TDB
 EC= 1.658972894380405E-02 QR= 9.842565408034613E-01 IN= 4.176751837366949E-03
 OM= 1.798997882563012E+02 W = 2.858589126517139E+02 Tp=  2462142.684540513903
 N = 9.843382729120017E-01 MA= 9.583234541296071E+01 TA= 9.771922931564085E+01
 A = 1.000860545971679E+00 AD= 1.017464551139897E+00 PR= 3.657279310444769E+02
2462240.083333333 = A.D. 2029-Apr-13 14:00:00.0000 TDB
 EC= 1.658615135794912E-02 QR= 9.842604650233783E-01 IN= 4.192863646154958E-03
 OM= 1.799959359447945E+02 W = 2.857627124120366E+02 Tp=  2462142.684021904133
 N = 9.843377575382314E-01 MA= 9.587381979794885E+01 TA= 9.776012976836853E+01
 A = 1.000860895321432E+00 AD= 1.017461325619486E+00 PR= 3.657281225301547E+02
2462240.125000000 = A.D. 2029-Apr-13 15:00:00.0000 TDB
 EC= 1.658257889756766E-02 QR= 9.842643286080860E-01 IN= 4.208952563090501E-03
 OM= 1.800913288501132E+02 W = 2.856670636102882E+02 Tp=  2462142.683304712642
 N = 9.843373254328112E-01 MA= 9.591549772495434E+01 TA= 9.780123261600430E+01
 A = 1.000861188227380E+00 AD= 1.017458047846674E+00 PR= 3.657282830778653E+02
2462240.166666667 = A.D. 2029-Apr-13 16:00:00.0000 TDB
 EC= 1.657901185015733E-02 QR= 9.842681312385337E-01 IN= 4.225017327349151E-03
 OM= 1.801859701571826E+02 W = 2.855719629162240E+02 Tp=  2462142.682388825808
 N = 9.843369766416062E-01 MA= 9.595737931969254E+01 TA= 9.784253800304542E+01
 A = 1.000861424658309E+00 AD= 1.017454718078084E+00 PR= 3.657284126704860E+02
2462240.208333333 = A.D. 2029-Apr-13 17:00:00.0000 TDB
 EC= 1.657545050379425E-02 QR= 9.842718725973526E-01 IN= 4.241056679100838E-03
 OM= 1.802798630538209E+02 W = 2.854774070067455E+02 Tp=  2462142.681274142582
 N = 9.843367112071260E-01 MA= 9.599946469781578E+01 TA= 9.788404606395015E+01
 A = 1.000861604585304E+00 AD= 1.017451336573256E+00 PR= 3.657285112921569E+02
2462240.250000000 = A.D. 2029-Apr-13 18:00:00.0000 TDB
 EC= 1.657189514712401E-02 QR= 9.842755523688699E-01 IN= 4.257069359618749E-03
 OM= 1.803730107291133E+02 W = 2.853833925675882E+02 Tp=  2462142.679960568435
 N = 9.843365291685179E-01 MA= 9.604175396485115E+01 TA= 9.792575692307541E+01
 A = 1.000861727981753E+00 AD= 1.017447903594636E+00 PR= 3.657285789282825E+02
```

```
2462240.291666667 = A.D. 2029-Apr-13 19:00:00.0000 TDB
 EC= 1.656834606934795E-02 QR= 9.842791702391316E-01 IN= 4.273054111327722E-03
 OM= 1.804654163716150E+02 W = 2.852899162951952E+02 Tp=  2462142.678448021878
 N = 9.843364305615546E-01 MA= 9.608424721612450E+01 TA= 9.796767069460064E+01
 A = 1.000861794823354E+00 AD= 1.017444419407576E+00 PR= 3.657286155655373E+02
2462240.333333333 = A.D. 2029-Apr-13 20:00:00.0000 TDB
 EC= 1.656480356021319E-02 QR= 9.842827258959130E-01 IN= 4.289009677843255E-03
 OM= 1.805570831672932E+02 W = 2.851969748988215E+02 Tp=  2462142.676736431196
 N = 9.843364154186321E-01 MA= 9.612694453671408E+01 TA= 9.800978748248079E+01
 A = 1.000861805088118E+00 AD= 1.017440884280324E+00 PR= 3.657286211918658E+02
2462240.375000000 = A.D. 2029-Apr-13 21:00:00.0000 TDB
 EC= 1.656126790999936E-02 QR= 9.842862190287444E-01 IN= 4.304934804101903E-03
 OM= 1.806480142983887E+02 W = 2.851045651017574E+02 Tp=  2462142.674825733062
 N = 9.843364837687547E-01 MA= 9.616984600136679E+01 TA= 9.805210738036260E+01
 A = 1.000861758756382E+00 AD= 1.017437298484019E+00 PR= 3.657285957964888E+02
2462240.416666667 = A.D. 2029-Apr-13 22:00:00.0000 TDB
 EC= 1.655773940950647E-02 QR= 9.842896493289245E-01 IN= 4.320828236359021E-03
 OM= 1.807382129414410E+02 W = 2.850126836433579E+02 Tp=  2462142.672715876717
 N = 9.843366356375279E-01 MA= 9.621295167444293E+01 TA= 9.809463047152887E+01
 A = 1.000861655810807E+00 AD= 1.017433662292689E+00 PR= 3.657285393699055E+02
2462240.458333333 = A.D. 2029-Apr-13 23:00:00.0000 TDB
 EC= 1.655421835004232E-02 QR= 9.842930164895424E-01 IN= 4.336688722296129E-03
 OM= 1.808276822658778E+02 W = 2.849213272805226E+02 Tp=  2462142.670406820718
 N = 9.843368710471506E-01 MA= 9.625626160984777E+01 TA= 9.813735682882988E+01
 A = 1.000861496236390E+00 AD= 1.017429975983237E+00 PR= 3.657284519038967E+02
2462240.500000000 = A.D. 2029-Apr-14 00:00:00.0000 TDB
 EC= 1.655070502341066E-02 QR= 9.842963202054928E-01 IN= 4.352515011077305E-03
 OM= 1.809164254323657E+02 W = 2.848304927894075E+02 Tp=  2462142.667898533400
 N = 9.843371900164063E-01 MA= 9.629977585096897E+01 TA= 9.818028651462053E+01
 A = 1.000861280020465E+00 AD= 1.017426239835437E+00 PR= 3.657283333915279E+02
$$EOE
*******************************************************************************

TIME

  Barycentric Dynamical Time ("TDB" or T_eph) output was requested. This
continuous relativistic coordinate time is equivalent to the relativistic
proper time of a clock at rest in a reference frame comoving with the
solar system barycenter but outside the system's gravity well. It is the
independent variable in the solar system relativistic equations of motion.

  TDB runs at a uniform rate of one SI second per second and is independent
of irregularities in Earth's rotation.

  Calendar dates prior to 1582-Oct-15 are in the Julian calendar system.
Later calendar dates are in the Gregorian system.

REFERENCE FRAME AND COORDINATES

  Ecliptic at the standard reference epoch

    Reference epoch: J2000.0
    X-Y plane: adopted Earth orbital plane at the reference epoch
               Note: IAU76 obliquity of 84381.448 arcseconds wrt ICRF X-Y plane
```

```
    X-axis   : ICRF
    Z-axis   : perpendicular to the X-Y plane in the directional (+ or -) sense
                 of Earth's north pole at the reference epoch.

  Symbol meaning [1 au= 149597870.700 km, 1 day= 86400.0 s]:

    JDTDB    Julian Day Number, Barycentric Dynamical Time
      EC     Eccentricity, e
      QR     Periapsis distance, q (au)
      IN     Inclination w.r.t X-Y plane, i (degrees)
      OM     Longitude of Ascending Node, OMEGA, (degrees)
      W      Argument of Perifocus, w (degrees)
      Tp     Time of periapsis (Julian Day Number)
      N      Mean motion, n (degrees/day)
      MA     Mean anomaly, M (degrees)
      TA     True anomaly, nu (degrees)
      A      Semi-major axis, a (au)
      AD     Apoapsis distance (au)
      PR     Sidereal orbit period (day)

ABERRATIONS AND CORRECTIONS

 Geometric osculating elements have NO corrections or aberrations applied.

Computations by ...

    Solar System Dynamics Group, Horizons On-Line Ephemeris System
    4800 Oak Grove Drive, Jet Propulsion Laboratory
    Pasadena, CA  91109   USA

    General site: https://ssd.jpl.nasa.gov/
    Mailing list: https://ssd.jpl.nasa.gov/email_list.html
    System news : https://ssd.jpl.nasa.gov/horizons/news.html
    User Guide  : https://ssd.jpl.nasa.gov/horizons/manual.html
    Connect     : browser        https://ssd.jpl.nasa.gov/horizons/app.html#/x
                  API            https://ssd-api.jpl.nasa.gov/doc/horizons.html
                  command-line   telnet ssd.jpl.nasa.gov 6775
                  e-mail/batch   https://ssd.jpl.nasa.gov/ftp/ssd/hrzn_batch.txt
                  scripts        https://ssd.jpl.nasa.gov/ftp/ssd/SCRIPTS
    Author      : Jon.D.Giorgini@jpl.nasa.gov
*******************************************************************************
```

## Appendix A2 (calculations of ephemeris of Apophis's orbit for 2029-Apr-13 with time-step 1 hour, using on-line service NASA [9]).

Let us we calculate ephemeris of Apophis's orbit on 2029-Apr-13 with time-step 1 hour, using on-line service NASA [9]:

```
*******************************************************************************
JPL/HORIZONS                 99942 Apophis (2004 MN4)        2022-Jan-03 03:09:20
Rec #:   99942 (+COV) Soln.date: 2021-Jun-29_11:09:44   # obs: 7300 (2004-2021)

IAU76/J2000 helio. ecliptic osc. elements (au, days, deg., period=Julian yrs):

  EPOCH=  2459215.5 ! 2021-Jan-01.00 (TDB)         Residual RMS= .17997
   EC= .1915216892753218   QR= .745827047918679    TP= 2459101.0394224189
   OM= 204.038927353109    W=  126.6520516484658   IN= 3.336751317384612
   A= .9225071817327519    MA= 127.3225632492728   ADIST= 1.099187315546825
   PER= .88606             N= 1.112370411          ANGMOM= .016216289
   DAN= 1.00339            DDN= .79749             L= 330.7375495
   B= 2.676454             MOID= .0003822          TP= 2020-Sep-08.5394224189

Asteroid physical parameters (km, seconds, rotational period in hours):
   GM= n.a.                RAD= .170               ROTPER= 30.4
   H= 19.7                 G= .250                 B-V= n.a.
                           ALBEDO= .230            STYP= Sq

Asteroid non-gravitational force model (AMRAT= m^2/kg;A1,A2,A3=au/d^2;R0=au):
   AMRAT=  0.
   A1= 4.999999873689E-13  A2= -2.901085508711E-14 A3= 0.
 Non-standard or simulated/proxy model:
   ALN=  1.            NK=  0.       NM=  2.       NN=  5.093    R0=  1.

ASTEROID comments:
1: soln ref.= JPL#216, PHA  OCC=0       radar(20 delay,30 Dop.)
2: source=ORB
*******************************************************************************

*******************************************************************************
Ephemeris / WWW_USER Mon Jan  3 03:09:21 2022 Pasadena, USA      / Horizons
*******************************************************************************
Target body name: 99942 Apophis (2004 MN4)        {source: JPL#216}
Center body name: Sun (10)                        {source: DE441}
Center-site name: BODY CENTER
*******************************************************************************
Start time      : A.D. 2029-Apr-13 00:00:00.0000 TDB
Stop  time      : A.D. 2029-Apr-14 00:00:00.0000 TDB
Step-size       : 60 minutes
*******************************************************************************
Center geodetic : 0.00000000,0.00000000,0.0000000 {E-lon(deg),Lat(deg),Alt(km)}
Center cylindric: 0.00000000,0.00000000,0.0000000 {E-lon(deg),Dxy(km),Dz(km)}
Center radii    : 696000.0 x 696000.0 x 696000.0 k{Equator, meridian, pole}
Keplerian GM    : 2.9591220828411951E-04 au^3/d^2
Small perturbers: Yes                             {source: SB441-N16}
Output units    : AU-D, deg, Julian Day Number (Tp)
Output type     : GEOMETRIC osculating elements
Output format   : 10
Reference frame : Ecliptic of J2000.0
*******************************************************************************
Initial IAU76/J2000 heliocentric ecliptic osculating elements (au, days, deg.):
  EPOCH=  2459215.5 ! 2021-Jan-01.00 (TDB)         Residual RMS= .17997
```

```
   EC= .1915216892753218   QR= .745827047918679    TP= 2459101.0394224189
   OM= 204.038927353109    W=  126.6520516484658   IN= 3.336751317384612
  Equivalent ICRF heliocentric cartesian coordinates (au, au/d):
   X=-4.098530839654745E-01  Y= 9.070198867091532E-01  Z= 3.267919472096975E-01
  VX=-1.507552412128397E-02 VY=-3.713559925690538E-03 VZ=-1.761867988770488E-03
Asteroid physical parameters (km, seconds, rotational period in hours):
   GM= n.a.                RAD= .170               ROTPER= 30.4
   H= 19.7                 G= .250                 B-V= n.a.
                           ALBEDO= .230            STYP= Sq
Asteroid non-gravitational force model (AMRAT= m^2/kg;A1,A2,A3=au/d^2;R0=au):
   AMRAT=  0.
   A1= 4.999999873689E-13  A2= -2.901085508711E-14 A3= 0.
 Non-standard or simulated/proxy model:
   ALN=  1.            NK=  0.       NM=  2.       NN=  5.093    R0=  1.
*******************************************************************************
JDTDB
   EC     QR   IN
   OM     W    Tp
   N      MA   TA
   A      AD   PR
*******************************************************************************
$$SOE
2462239.500000000 = A.D. 2029-Apr-13 00:00:00.0000 TDB
 EC= 1.956772960803659E-01 QR= 7.410939273068958E-01 IN= 3.422705924904834E+00
 OM= 2.037857101141501E+02 W = 1.266230776136077E+02 Tp=  2462336.385685577523
 N = 1.114396357374862E+00 MA= 2.520309449105974E+02 TA= 2.325777769979668E+02
 A = 9.213887954366933E-01 AD= 1.101683663566491E+00 PR= 3.230448463130635E+02
2462239.541666667 = A.D. 2029-Apr-13 01:00:00.0000 TDB
 EC= 1.958736039480826E-01 QR= 7.409026142704516E-01 IN= 3.426403089170472E+00
 OM= 2.037848681882018E+02 W = 1.266105144407129E+02 Tp=  2462336.349975794088
 N = 1.114419905371442E+00 MA= 2.521148933029008E+02 TA= 2.326294052514206E+02
 A = 9.213758159266000E-01 AD= 1.101849017582748E+00 PR= 3.230380202873441E+02
2462239.583333333 = A.D. 2029-Apr-13 02:00:00.0000 TDB
 EC= 1.960880203992830E-01 QR= 7.406953228411089E-01 IN= 3.430447017475155E+00
 OM= 2.037839944871199E+02 W = 1.265961346865145E+02 Tp=  2462336.310623253230
 N = 1.114441871924059E+00 MA= 2.522030579558207E+02 TA= 2.326828889550684E+02
 A = 9.213637085107175E-01 AD= 1.102032094180326E+00 PR= 3.230316529461227E+02
2462239.625000000 = A.D. 2029-Apr-13 03:00:00.0000 TDB
 EC= 1.963231895454603E-01 QR= 7.404699781969480E-01 IN= 3.434889061003230E+00
 OM= 2.037830866428652E+02 W = 1.265795732811472E+02 Tp=  2462336.267039036844
 N = 1.114461441337462E+00 MA= 2.522961738810482E+02 TA= 2.327385954305316E+02
 A = 9.213529226731783E-01 AD= 1.102235867149409E+00 PR= 3.230259806637769E+02
2462239.666666667 = A.D. 2029-Apr-13 04:00:00.0000 TDB
 EC= 1.965822968063714E-01 QR= 7.402241411931583E-01 IN= 3.439791228123944E+00
 OM= 2.037821420424757E+02 W = 1.265603674092132E+02 Tp=  2462336.218500365503
 N = 1.114477493279086E+00 MA= 2.523951544078703E+02 TA= 2.327969899782418E+02
 A = 9.213440757537795E-01 AD= 1.102464010314401E+00 PR= 3.230213280851329E+02
2462239.708333333 = A.D. 2029-Apr-13 05:00:00.0000 TDB
 EC= 1.968692144666637E-01 QR= 7.399549296089983E-01 IN= 3.445229088878626E+00
 OM= 2.037811578030879E+02 W = 1.265379230133518E+02 Tp=  2462336.164110430982
 N = 1.114488476910507E+00 MA= 2.525011478932161E+02 TA= 2.328586693981728E+02
 A = 9.213380223217510E-01 AD= 1.102721115034504E+00 PR= 3.230181446092313E+02
2462239.750000000 = A.D. 2029-Apr-13 06:00:00.0000 TDB
 EC= 1.971886965703383E-01 QR= 7.396589206349231E-01 IN= 3.451295677020576E+00
```

```
 OM= 2.037801307485486E+02 W = 1.265114670035224E+02 Tp=  2462336.102742932271
 N = 1.114492224144517E+00 MA= 2.526156172270373E+02 TA= 2.329244098036765E+02
 A = 9.213359571234890E-01 AD= 1.103012993612055E+00 PR= 3.230170585320465E+02
2462239.791666667 = A.D. 2029-Apr-13 07:00:00.0000 TDB
 EC= 1.975466436351619E-01 QR= 7.393320309647277E-01 IN= 3.458106808969660E+00
 OM= 2.037790573916638E+02 W = 1.264799777174200E+02 Tp=  2462336.032964078244
 N = 1.114485667206200E+00 MA= 2.527404534413882E+02 TA= 2.329952361793646E+02
 A = 9.213395708305668E-01 AD= 1.103347110696406E+00 PR= 3.230189589628821E+02
2462239.833333333 = A.D. 2029-Apr-13 08:00:00.0000 TDB
 EC= 1.979504678515695E-01 QR= 7.389693717672423E-01 IN= 3.465808449878000E+00
 OM= 2.037779339296509E+02 W = 1.264420814878885E+02 Tp=  2462335.952920691110
 N = 1.114464402112836E+00 MA= 2.528781415440215E+02 TA= 2.330725258203302E+02
 A = 9.213512908458199E-01 AD= 1.103733209924397E+00 PR= 3.230251224870896E+02
2462239.875000000 = A.D. 2029-Apr-13 09:00:00.0000 TDB
 EC= 1.984096055559796E-01 QR= 7.385650794576855E-01 IN= 3.474587091934992E+00
 OM= 2.037767562668395E+02 W = 1.263958949565403E+02 Tp=  2462335.860176010989
 N = 1.114421997298948E+00 MA= 2.530320084388343E+02 TA= 2.331581660007515E+02
 A = 9.213746628912030E-01 AD= 1.104184246324720E+00 PR= 3.230374138993496E+02
2462239.916666667 = A.D. 2029-Apr-13 10:00:00.0000 TDB
 EC= 1.989362498164560E-01 QR= 7.381121336323518E-01 IN= 3.484684652306584E+00
 OM= 2.037755200907482E+02 W = 1.263387778753327E+02 Tp=  2462335.751462555025
 N = 1.114348867672848E+00 MA= 2.532066037181345E+02 TA= 2.332548011063894E+02
 A = 9.214149728573183E-01 AD= 1.104717812082285E+00 PR= 3.230586133692642E+02
2462239.958333333 = A.D. 2029-Apr-13 11:00:00.0000 TDB
 EC= 1.995464188859456E-01 QR= 7.376022006700432E-01 IN= 3.496420292162982E+00
 OM= 2.037742210507290E+02 W = 1.262669330663570E+02 Tp=  2462335.622299195733
 N = 1.114230381526754E+00 MA= 2.534083028191247E+02 TA= 2.333662325160224E+02
 A = 9.214802932650563E-01 AD= 1.105358385860070E+00 PR= 3.230929670996014E+02
2462240.000000000 = A.D. 2029-Apr-13 12:00:00.0000 TDB
 EC= 2.002615482587268E-01 QR= 7.370256101904250E-01 IN= 3.510223053658173E+00
 OM= 2.037728551335316E+02 W = 1.261747349447588E+02 Tp=  2462335.466377822217
 N = 1.114043558247903E+00 MA= 2.536462967581187E+02 TA= 2.334980897412133E+02
 A = 9.215833108758205E-01 AD= 1.106141011561216E+00 PR= 3.231471492606493E+02
2462240.041666667 = A.D. 2029-Apr-13 13:00:00.0000 TDB
 EC= 2.011109162734236E-01 QR= 7.363717467102028E-01 IN= 3.526681725123311E+00
 OM= 2.037714194213280E+02 W = 1.260535536101935E+02 Tp=  2462335.274542454164
 N = 1.113751055526532E+00 MA= 2.539342840708890E+02 TA= 2.336590056508976E+02
 A = 9.217446598159171E-01 AD= 1.107117572921632E+00 PR= 3.232320168978947E+02
2462240.083333333 = A.D. 2029-Apr-13 14:00:00.0000 TDB
 EC= 2.021354201504373E-01 QR= 7.356305990390459E-01 IN= 3.546622483557559E+00
 OM= 2.037699136082317E+02 W = 1.258895906000703E+02 Tp=  2462335.033011960331
 N = 1.113289655479321E+00 MA= 2.542935049934511E+02 TA= 2.338627793481402E+02
 A = 9.219993187036186E-01 AD= 1.108368038368191E+00 PR= 3.233659795796844E+02
2462240.125000000 = A.D. 2029-Apr-13 15:00:00.0000 TDB
 EC= 2.033935549291969E-01 QR= 7.347974788125944E-01 IN= 3.571231044651094E+00
 OM= 2.037683430638220E+02 W = 1.256596538028371E+02 Tp=  2462334.720117666293
 N = 1.112546857468001E+00 MA= 2.547584991084221E+02 TA= 2.341325984936747E+02
 A = 9.224096583191528E-01 AD= 1.110021837825711E+00 PR= 3.235818766495003E+02
2462240.166666667 = A.D. 2029-Apr-13 16:00:00.0000 TDB
 EC= 2.049708987349042E-01 QR= 7.338866548460146E-01 IN= 3.602241697273716E+00
 OM= 2.037667251541358E+02 W = 1.253223056031146E+02 Tp=  2462334.299925089348
 N = 1.111309756114471E+00 MA= 2.553887915397005E+02 TA= 2.345098854873661E+02
 A = 9.230940775352904E-01 AD= 1.112301500224566E+00 PR= 3.239420854710089E+02
2462240.208333333 = A.D. 2029-Apr-13 17:00:00.0000 TDB
```

```
 EC= 2.069944794632238E-01 QR= 7.329715974481782E-01 IN= 3.642198892971934E+00
 OM= 2.037651026116047E+02 W = 1.247976673606855E+02 Tp=  2462333.708786398638
 N = 1.109143327189900E+00 MA= 2.562945963934458E+02 TA= 2.350744811128568E+02
 A = 9.242957059769752E-01 AD= 1.115619814505772E+00 PR= 3.245748238075666E+02
2462240.250000000 = A.D. 2029-Apr-13 18:00:00.0000 TDB
 EC= 2.096497221144306E-01 QR= 7.323117115945276E-01 IN= 3.694623090440989E+00
 OM= 2.037635720655625E+02 W = 1.239163799965544E+02 Tp=  2462332.824333595112
 N = 1.105069315037514E+00 MA= 2.576989445839413E+02 TA= 2.359956591793809E+02
 A = 9.265660202634294E-01 AD= 1.120820328932331E+00 PR= 3.257714200378271E+02
2462240.291666667 = A.D. 2029-Apr-13 19:00:00.0000 TDB
 EC= 2.131708895072920E-01 QR= 7.327902155349251E-01 IN= 3.762756014150797E+00
 OM= 2.037623397909467E+02 W = 1.222747801011692E+02 Tp=  2462331.386491293553
 N = 1.096617595518388E+00 MA= 2.601038124533294E+02 TA= 2.376769052016843E+02
 A = 9.313206714937833E-01 AD= 1.129851127452641E+00 PR= 3.282821664281454E+02
2462240.333333333 = A.D. 2029-Apr-13 20:00:00.0000 TDB
 EC= 2.176301241225753E-01 QR= 7.375468534604454E-01 IN= 3.838284830392882E+00
 OM= 2.037617601493688E+02 W = 1.187855723846093E+02 Tp=  2462328.780793476384
 N = 1.076806906291594E+00 MA= 2.647591640743967E+02 TA= 2.412052386146776E+02
 A = 9.427086550760783E-01 AD= 1.147870456691711E+00 PR= 3.343217784884023E+02
2462240.375000000 = A.D. 2029-Apr-13 21:00:00.0000 TDB
 EC= 2.219691463239375E-01 QR= 7.569391927883349E-01 IN= 3.836646932616203E+00
 OM= 2.037616042070039E+02 W = 1.105694250243232E+02 Tp=  2462323.563036841806
 N = 1.027088376747083E+00 MA= 2.745585342752113E+02 TA= 2.494604536715505E+02
 A = 9.728909711124271E-01 AD= 1.188842749436519E+00 PR= 3.505053782617663E+02
2462240.416666667 = A.D. 2029-Apr-13 22:00:00.0000 TDB
 EC= 2.237750039267324E-01 QR= 8.025553929370386E-01 IN= 3.518396682990265E+00
 OM= 2.037567403095802E+02 W = 9.538422382170093E+01 Tp=  2462314.648721897043
 N = 9.375036850743244E-01 MA= 2.904071746709191E+02 TA= 2.646903320360903E+02
 A = 1.033921088597984E+00 AD= 1.265286784258929E+00 PR= 3.839984905994899E+02
2462240.458333333 = A.D. 2029-Apr-13 23:00:00.0000 TDB
 EC= 2.214896295921848E-01 QR= 8.439272695160072E-01 IN= 3.052071191810894E+00
 OM= 2.037423515942073E+02 W = 8.286358299982773E+01 Tp=  2462307.189766349271
 N = 8.732591534656736E-01 MA= 3.017261652950475E+02 TA= 2.772663593444736E+02
 A = 1.084028295055240E+00 AD= 1.324129320594472E+00 PR= 4.122487563643397E+02
2462240.500000000 = A.D. 2029-Apr-14 00:00:00.0000 TDB
 EC= 2.165603148889914E-01 QR= 8.643536451253669E-01 IN= 2.766571576170378E+00
 OM= 2.037270153347350E+02 W = 7.716249709496233E+01 Tp=  2462303.788095762953
 N = 8.505017715551982E-01 MA= 3.061733624353312E+02 TA= 2.830244753443160E+02
 A = 1.103280394843533E+00 AD= 1.342207144561699E+00 PR= 4.232795416072049E+02
$$EOE
*******************************************************************************
```

TIME

Barycentric Dynamical Time ("TDB" or T_eph) output was requested. This continuous relativistic coordinate time is equivalent to the relativistic proper time of a clock at rest in a reference frame comoving with the solar system barycenter but outside the system's gravity well. It is the independent variable in the solar system relativistic equations of motion.

TDB runs at a uniform rate of one SI second per second and is independent of irregularities in Earth's rotation.

Calendar dates prior to 1582-Oct-15 are in the Julian calendar system.

```
Later calendar dates are in the Gregorian system.

REFERENCE FRAME AND COORDINATES

  Ecliptic at the standard reference epoch

    Reference epoch: J2000.0
    X-Y plane: adopted Earth orbital plane at the reference epoch
               Note: IAU76 obliquity of 84381.448 arcseconds wrt ICRF X-Y plane
    X-axis   : ICRF
    Z-axis   : perpendicular to the X-Y plane in the directional (+ or -) sense
               of Earth's north pole at the reference epoch.

  Symbol meaning [1 au= 149597870.700 km, 1 day= 86400.0 s]:

    JDTDB    Julian Day Number, Barycentric Dynamical Time
      EC     Eccentricity, e
      QR     Periapsis distance, q (au)
      IN     Inclination w.r.t X-Y plane, i (degrees)
      OM     Longitude of Ascending Node, OMEGA, (degrees)
      W      Argument of Perifocus, w (degrees)
      Tp     Time of periapsis (Julian Day Number)
      N      Mean motion, n (degrees/day)
      MA     Mean anomaly, M (degrees)
      TA     True anomaly, nu (degrees)
      A      Semi-major axis, a (au)
      AD     Apoapsis distance (au)
      PR     Sidereal orbit period (day)

ABERRATIONS AND CORRECTIONS

 Geometric osculating elements have NO corrections or aberrations applied.

Computations by ...

    Solar System Dynamics Group, Horizons On-Line Ephemeris System
    4800 Oak Grove Drive, Jet Propulsion Laboratory
    Pasadena, CA  91109   USA

    General site: https://ssd.jpl.nasa.gov/
    Mailing list: https://ssd.jpl.nasa.gov/email_list.html
    System news : https://ssd.jpl.nasa.gov/horizons/news.html
    User Guide  : https://ssd.jpl.nasa.gov/horizons/manual.html
    Connect     : browser        https://ssd.jpl.nasa.gov/horizons/app.html#/x
                  API            https://ssd-api.jpl.nasa.gov/doc/horizons.html
                  command-line   telnet ssd.jpl.nasa.gov 6775
                  e-mail/batch   https://ssd.jpl.nasa.gov/ftp/ssd/hrzn_batch.txt
                  scripts        https://ssd.jpl.nasa.gov/ftp/ssd/SCRIPTS
    Author      : Jon.D.Giorgini@jpl.nasa.gov
*******************************************************************************
```

## Declarations

On behalf of all authors, the corresponding author states that there is no conflict of interest. The data for this paper are available by contacting the corresponding author. Remark regarding contributions of authors as below:

In this research, Dr. Sergey Ershkov is responsible for the main idea and general ansatz, suggested algorithm, simple algebraic calculating, results of the article and also is responsible for the obtaining approximate estimations.
Prof. Dmytro Leshchenko is responsible for theoretical investigations and deep survey in literature on the problem under consideration.
All authors agreed with results and conclusions of each other in Sections 1-3.